\documentclass[12pt]{mn2e}
\usepackage{graphicx}
\usepackage{epsf}
\usepackage{lscape}
\usepackage{longtable}
\usepackage{rotating}
\usepackage{color}

\newcommand{\apj}{ApJ}
\newcommand{\apjl}{ApJ}

\newcommand{\mnras}{MNRAS}

\begin{document}
\topmargin -0.5in %this is only for astro-ph, uncomment when submitting paper

\title[Probing $\sim L_{*}$ Lyman-break Galaxies at $z\approx 7$ in GOODS-South]{Probing $\sim L_{*}$ Lyman-break Galaxies at $z\approx 7$ in GOODS-South with WFC3 on HST}

\author[Stephen M. Wilkins, et al.\ ]  
{
Stephen M. Wilkins$^{1}$\thanks{E-mail: stephen.wilkins@astro.ox.ac.uk}, Andrew J. Bunker$^{1}$,
Richard S. Ellis$^{2}$, 
Daniel Stark$^{3}$, \newauthor Elizabeth R. Stanway$^{4}$, Kuenley Chiu$^{2}$, Silvio Lorenzoni$^{1}$, Matt J. Jarvis$^{5}$ \\
$^1$\,University of Oxford, Department of Physics, Denys Wilkinson Building, Keble Road, OX1 3RH, U.K. \\
$^{2}$\,California Institute of
Technology, Astrophysics Department, Mail Stop 169-327, Pasadena, CA~91109, U.S.A.\\
$^{3}$\,University of Cambridge, Institute of Astrophysics, 
Madingley Road, Cambridge, CB3\,0HA, U.K.\\
$^{4}$\,H H Wills Physics Laboratory, Tyndall Avenue, Bristol, BS8 1TL, U.K.\\ 
$^{5}$\,Centre for Astrophysics, Science \& Technology Research Institute, University of Hertfordshire, Hatfield, Herts, AL10 9AB, U.K.}
\maketitle

\begin{abstract}
We analyse recently acquired near-infrared {\em Hubble} Space Telescope imaging of the GOODS-South field to search for star forming galaxies at $z\approx 7.0$. By comparing WFC\,3 $0.98\mu$m $Y$-band images with ACS $z$-band ($0.85\mu$m) images, we identify objects with colours consistent with Lyman break galaxies at z$\simeq$6.4-7.4. This new data covers an area five times larger than that previously reported in the WFC3 imaging of the {\em Hubble} Ultra Deep Field, and affords a valuable constraint on the bright end of the luminosity function. Using additional imaging of the region in the ACS $B$, $V$ and $i$-bands from GOODS v2.0 and the WFC3 $J$-band we attempt to remove any low-redshift interlopers. Our selection criteria yields 6 candidates brighter than $Y_{\rm AB}=27.0$, of which all except one are detected in the ACS $z$-band imaging and are thus unlikely to be transients. Assuming all 6 candidates are at $z\approx 7$ this implies a surface density of objects brighter than $Y_{\rm AB}=27.0$ of 0.30$\pm$0.12\,arcmin$^{-2}$, a value significantly smaller than the prediction from $z\approx 6$ luminosity function. This suggests continued evolution of the bright end of the luminosity function between $z=6\to 7$, with number densities lower at higher redshift. 
\end{abstract} 

\begin{keywords}  
galaxies: evolution –- galaxies: formation –- galaxies: starburst –- galaxies: high-redshift –- ultraviolet: galaxies
\end{keywords} 

\section{Introduction}

In recent years our understanding of the high-redshift galaxy
population has rapidly expanded with the discovery of star-forming galaxies
within the first billion years ($z>5$), through the Lyman break
technique using broad-band imaging (e.g. Stanway, Bunker \& McMahon 2003; Dickinson et al.\ 2004) and
searches for Lyman-$\alpha$ emission with narrow-band filters (e.g., Ouchi et al.\ 2008, Ota et al.\ 2008),
and most recently gamma-ray bursts (e.g. Tanvir et al.\ 2009, Salvaterra et al. 2009).
At $z\sim 6$, the brightest ($z<26.5$) Lyman-break ``$i$-band drop-out" galaxies have been confirmed
spectroscopically (e.g. Bunker et al.\ 2003; Stanway et al.\ 2004ab; Dow-Hygelund et al.\ 2007; Vanzella et al.\ 2009) through their Lyman-$\alpha$ emission, confirming the validity of this photometric redshift
selection.

In recent weeks, the installation of the new WFC3 camera on the {\em Hubble} Space Telescope ({\em HST}),
which has an infrared channel with a  large field of view,
has enabled the Lyman break technique to be pushed to $z\sim 7-10$, revealing for
the first time significant numbers of galaxy candidates close to the reionization epoch
(Bunker et al.\ 2009; Bouwens et al.\ 2009; Oesch et al.\ 2009; McLure et al.\ 2009; Yan et al.\ 2009). However,
the first data to be released was the extremely deep single pointing on the Hubble Ultra Deep Field
($\approx 4\,{\rm arcmin}^{2}$), reaching objects as faint at mag$_{AB}=28.5$ ($6\,\sigma$)
in $Y$-, $J$- and $H$-bands. 
To probe the rare galaxies at the bright end of the luminosity function at $z\sim 7-8$ requires
a larger field of view. There have been some shallower wider area searches
using ground based observations (e.g. Hickey et al. 2009, Ouchi et al. 2009, Castellano et al. 2009), but
the depths probed are typically $Y_{AB}<26$ (equivalent to $L_{UV}>2\,L^*$)
and the numbers of robust candidates are
small. Increasing the number of $z\approx 7$ candidates over a wide range of magnitudes is
critical to exploring the shape of the luminosity function. This is particularly important as there is strong evidence for
evolution of luminosity function, with suggestions that at high redshift there is a larger
relative contribution to the stellar mass and UV luminosity density from sub-$L_{*}$ systems.
At $z\sim 7$ we are close to the epoch of reionization, and an open question is the mechanism by which
reionization is achieved; if we are to address the contribution of the UV from star-forming galaxies,
then quantifying the luminosity function is vital (along with the escape fraction of ionizing photons
from galaxies, and the hardness of the UV spectral slope).

In this paper we present first results from {\em HST}/WFC3 imaging of some of
the GOODS-South field, covering an area 5 times that of the WFC3 images
of the {\em Hubble Ultra Deep Field}, and reaching $Y_{AB}=27.0$ ($6\,\sigma$),
probing luminosities around $L^*_{UV z\sim 6}$.
In conjunction with the GOODS v2.0 Advanced Camera for Surveys (ACS) images with $B,V,i,z$ bands (Giavalisco et al.\ 2004)
 we search for objects which are much brighter in the $Y$-band WFC3 filter at
$1\,\mu$m than the $z$-band $0.9\,\mu$m, and are undetected at shorter wavelength.
These ``$z$-drops'' are candidate $z\sim 7$ Lyman break galaxies, and 
the greater area of this new dataset (compared with the Hubble Ultra Deep Field
WFC3 images) means that we are likely to find brighter objects more amenable to future spectroscopic confirmation.

The structure of this paper is as follows. In Section 2 the imaging
data and the construction of catalogues  is
described. In Section 3 we describe our candidate selection and
discuss the observed  surface density of $z\approx 7$
galaxies, comparing with the expected number from a range of
luminosity functions. Our conclusions are presented in Section 4. Throughout we
adopt the standard ``concordance'' cosmology of $\Omega_{M}=0.3$,
$\Omega_{\Lambda}=0.7$, and use $h_{70}=H_{0}/70 {\rm
  kms^{-1}\,Mpc^{-1}}$. All magnitudes are on the AB system (Oke \&
  Gunn 1983).

\section{HST Observations and Data Reduction}

We use the {\em Hubble} Space Telescope images of GOODS-South obtained
with WFC\,3 under the Early Release Science program GO/DD\#11359
(P.I.\ R.~O'Connell). This will ultimately cover 10 pointings with
the infrared channel within
the GOODS-South field, along with UV images of the same area.
We focus here on the first 6 WFC\,3 pointings to be released (data taken over the period 17-27 September 2009 UT). The
observations were split into visits, and each pointing was imaged
for 2 orbits in each filter during a single visit.

Each pointing was imaged in 3 filters (F098M ``$Y$-band", F125W ``$J$-band"
and F160W ``$H$-band"), and each filter observation comprised
two orbits (taken within the same {\em HST} visit) which were split into
three exposures. The near-IR channel of WFC\,3 has a $1014\times 1014$ Teledyne
HgCdTe detector
which can be read non-destructively. Each exposure involved a ``MULTIACCUM"
sequence of non-destructive reads (the SPARS100 pattern was used,
spacing reads by $\approx 100$\,s), enabling cosmic ray rejection through
``sampling up the ramp". The exposures comprised 9 or 10 non-destructive
read-outs, totalling 803--903\,s. The IRAF.STSDAS pipeline {\tt calwfc3} was used
to calculate the count rate and reject cosmic rays through gradient fitting,
as well as subtracting the zeroth read and flat-fielding. 
From each individual exposure in each filter, we subtracted the median stack
of the 36 exposures in the same filter (6 per pointing and 6 pointings), with bright objects
 excluded from the median using object masks created by the {\tt xdimsum} package. 
 This subtracted out hot pixels and eliminated quadrant pedestal offset effects, as well 
 as correcting for scattered light. We used MULTIDRIZZLE
(Koekemoer et al.\ 2002) to combine the 6 exposures per filter in each pointing,
taking account of the geometric distortions and mapping on to an output
pixel size of $0\farcs06$ from an original $0\farcs13\,{\rm pix}^{-1}$.
This was the same scale as we used in our analysis of the Hubble Ultra Deep Field
WFC\,3 images (Bunker et al.\ 2009) and corresponds to a $2\times 2$ block-averaging of the 
GOODSv2.0 ACS drizzled images. The 6 exposures
per pointing were dithered by 10\,arcsec, so that bad pixels did not overlap
and to facilitate subtraction of any residual background (see above). We found that it was
necessary to introduce small corrections to the pointing information in the header
to accurately align the exposures in MULTIDRIZZLE (and prevent real objects
being rejected by the algorithm). There was a slight discrepancy between
the position angle information in the headers and the GOODS v2.0 mosaic,
which we corrected for by introducing a $0.3^{\circ}$ relative rotation.
The total exposure time in each filter per pointing was 5218\,s, and the 6
pointings were reduced separately and all registered to the GOODS v2.0
ACS mosaics rebinned by $2\times 2$ pixels. Each pointing surveyed a region of $3.3\,{\rm arcmin}^{2}$
to maximum depth (i.e., where all 6 exposures overlapped), with the edge
regions covered by fewer exposures. In this analysis we only consider the deepest region
of uniform depth, covering a total of $20\,{\rm arcmin}^{2}$ over the 6 pointings.

The final frames had units of electrons/sec, and we take the standard ACS zeropoints for the UDF images. For WFC3,
we use the recent zeropoints reported on {\tt http://www.stsci.edu/hst/wfc3/phot\_zp\_lbn},
where the F098M $Y$-band has an AB magnitude zeropoint of 25.68 (such that a source of this brightness would have a count rate of 1 electron per second), and the F125W $J$-band zeropoint is 26.25. We note that the information in
the image headers is slightly different by 0.1-0.15\,mag, with zeropoints of $Y_{ZP}=25.58$ and $J_{ZP}=26.10$.

\begin{figure}
\centering
\includegraphics[width=18pc]{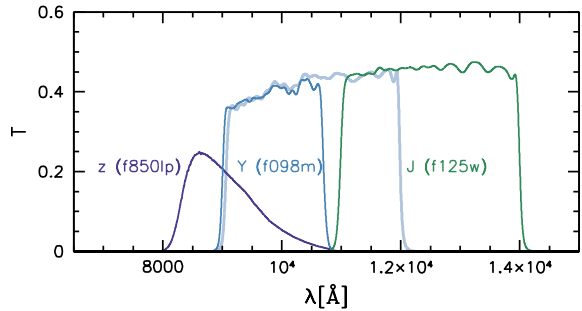} 
\caption{Transmission curves for the ACS $z'_{f850lp}$ and WFC3
  $Y_{f098m}$ and $J_{f125w}$ filters. Also shown is the curve for the
  $Y_{f105w}$ filter recently employed in studies of the HUDF.}
\end{figure}

\subsection{Construction of Catalogues}

Candidate selection for all objects in the field was performed
using version 2.5.0 of the SExtractor photometry package (Bertin
\& Arnouts 1996). For the $z$-drops, as we are searching specifically for objects
which are only securely detected in the WFC\,3 $Y$-band, with minimal flux in the
ACS images, fixed circular apertures $0\farcs6$ in diameter were
`trained' in the $Y$-image and the identified apertures used to
measure the flux at the same spatial location in the $z'$-band
image by running SExtractor in dual-image mode. This was repeated
for all other ACS and WFC\,3 filters. For object
identification, we adopted a limit of at least 5 contiguous pixels
above a threshold of $2\sigma$ per pixel (on the data drizzled to a
scale of 0\farcs06~pixel$^{-1}$). This cut enabled us to detect all
significant sources and a number of spurious detections close to
the noise limit, or due to diffraction spikes of stars. As high redshift galaxies in the rest-UV are
known to be compact (e.g., Ferguson et al.\ 2004, Bremer et al.\
2004, Bunker et al.\ 2004), we corrected the aperture magnitudes to approximate total
magnitudes with the aperture correction appropriate for that filter; the aperture
corrections were 0.23\, mag in $Y$-band and 0.25\,mag in $J$-band,
with the ACS bands having aperture corrections of $\approx$\,0.1\,mag.

The ``multidrizzle'' procedure resamples and interpolates pixels, resulting in
noise which is highly correlated. Rather than using the standard deviation of counts in the drizzled
images to determine the noise level, we instead look at the individual reduced exposures (after flatfielding
and background subtraction), and a combination using integer-pixel shifts which preserves the noise
characteristics. The noise in an individual $Y$-band exposure is  $0.026\,e^{-}\,s^{-1}$,
and in a combined pointing it is $0.011\,e^{-}\,s^{-1}$. We impose a brightness
cut of $Y_{AB}<27.0$ (after applying the aperture correction) corresponding to a $>6\,\sigma$ detection.
Working at this secure signal-to-noise ensures purity of the sample and robustness of the broad-band
colours. Our limiting magnitude of $Y_{AB}=27.0$ is equivalent to a star formation rate of $4.5\,{\rm
  M_{\odot}\,yr^{-1}}$ at $z=7$, using the conversion from UV continuum luminosity of Madau, Pozzetti \& Dickinson (1998) and
  assuming a Salpeter (1955) initial mass function and no correction for dust attenuation.

\begin{figure}
\centering
\includegraphics[width=18pc]{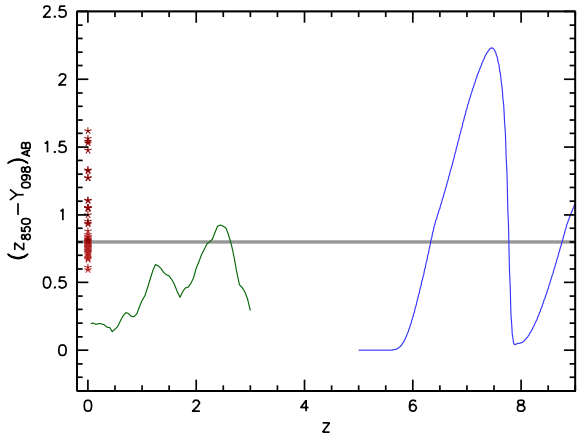} 
\caption{The evolution of the $(z'-Y)$ colour of starbursting
 galaxies, an instantaneous burst at $z=10$ and low-mass L and T dwarf
 stars. The blue line shows the colour evolution of a galaxy with
  ultraviolet spectral slopes of $\beta=-2.0$ (where
  $f_{\lambda}=\lambda^{\beta}$) damped below Lyman-$\alpha$ by a
  factor of $100$. The green line shows the colour evolution of an
  evolving stellar population that formed instantaneously at $z=10$ and
  the red stars denote the observed colours of L and T dwarf stars
  (Chiu et al. 2006, Golimowski et al. 2004, Knapp et al. 2004). The grey horizonal line denotes our $(z'-Y)_{\rm AB}=0.8$ colour selection criteria.}
\end{figure}

\begin{figure}
\centering
\includegraphics[width=18pc]{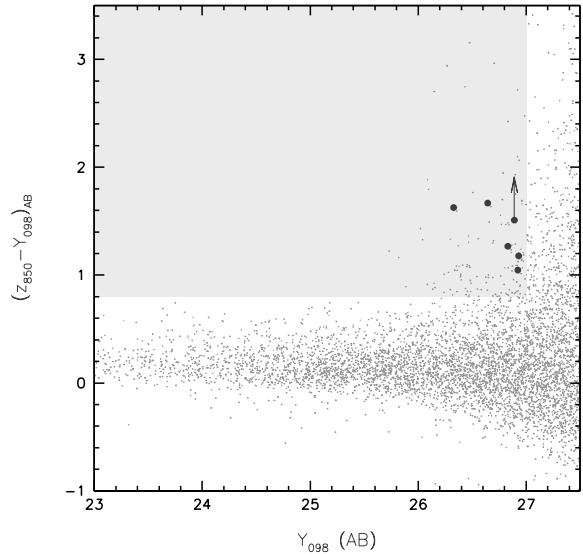} 
\caption{The $(z'-Y)$ colours of all objects in our catalogue (dots) as
  a function of $Y$-band magnitude. The grey shaded area denotes our $Y_{\rm AB}$-band magnitude and $(z'-Y)_{\rm AB}$ colour criteria. Objects meeting our full (undetected in the ACS $B,V,i'$ imaging and consistent $(Y-J)_{AB}$ colour) selection criteria are marked by the larger circles.}
\end{figure}

\begin{table*}
\begin{tabular}{c|c|c|c|clclclc}
ID &  RA Declination J2000 & $z_{\rm AB}$ & $Y_{\rm AB}$ & $J_{\rm AB}$ & $(z-Y)_{AB}$ & $(Y-J)_{AB}$ & SFR$_{z=7}$ (${\rm M_{\odot}\,yr^{-1}}$)\\%^{\dagger}$\\
\hline\hline
1 & 03:32:15.997  -27:43:01.44 & 27.95$\pm$ 0.44 & 26.32$\pm$ 0.07 & 26.48$\pm$ 0.05 & 1.62$\pm$0.44 & -0.16$\pm$0.09 & 8.6\\
2 &  03:32:25.285  -27:43:24.25  & 28.31$\pm$ 0.56 & 26.64$\pm$ 0.09 & 26.53$\pm$ 0.06 & 1.66$\pm$0.57 & 0.11$\pm$0.11 & 6.3\\
3 &  03:32:29.693  -27:40:49.88 & 28.09$\pm$ 0.47 & 26.82$\pm$ 0.11 & 27.30$\pm$ 0.12 & 1.26$\pm$0.48 & -0.48$\pm$0.17 & 5.3\\
4 &  03:32:29.541  -27:42:04.49  & $>$28.4 (2$\sigma$) & 26.89$\pm$ 0.12 & 26.82$\pm$ 0.08 & $>$1.51 (2$\sigma$) & 0.07$\pm$0.14 & 5.0 \\
5 &  03:32:37.833  -27:43:36.55 & 27.96$\pm$ 0.41 & 26.92$\pm$ 0.12 & 27.22$\pm$ 0.11 & 1.04$\pm$0.43 & -0.30$\pm$0.17 & 4.8 \\
6 &  03:32:24.094  -27:42:13.85& 28.10$\pm$ 0.47 & 26.93$\pm$ 0.12 & 26.49$\pm$ 0.05 & 1.17$\pm$0.49 & 0.44$\pm$0.13 & 4.8\\
\end{tabular}
%$^{\dagger}$\,The star formation rate, assuming the object is a galaxy at $z=7$.
\caption{$z$-band drop out candidate $z\approx 7$ galaxies meeting our selection
  criteria.  The star formation rate (SFR) has been derived using the conversion from UV luminosity
density from Madau, Pozzetti \& Dickinson (1998) and assuming the galaxies lie at $z=7.0$.}
\label{tab}
\end{table*}

\begin{figure*}
\centering
\includegraphics[width=32pc]{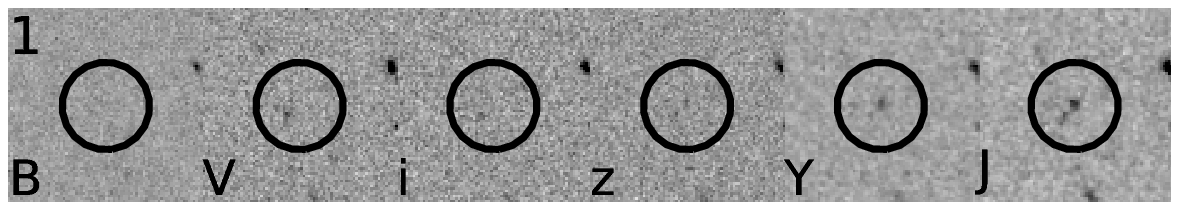} 
\includegraphics[width=32pc]{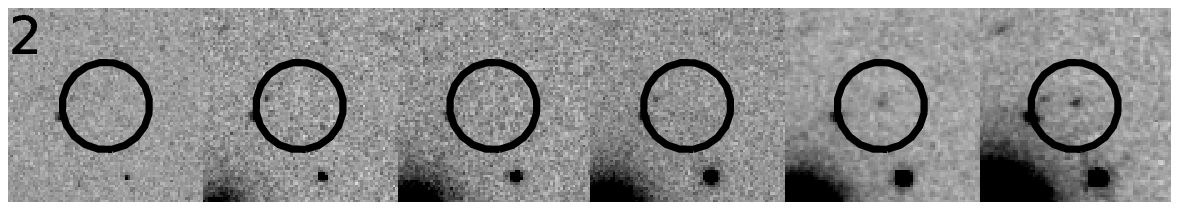} 
\includegraphics[width=32pc]{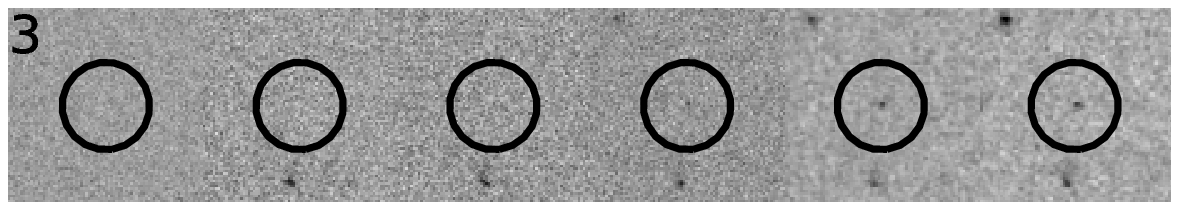} 
\includegraphics[width=32pc]{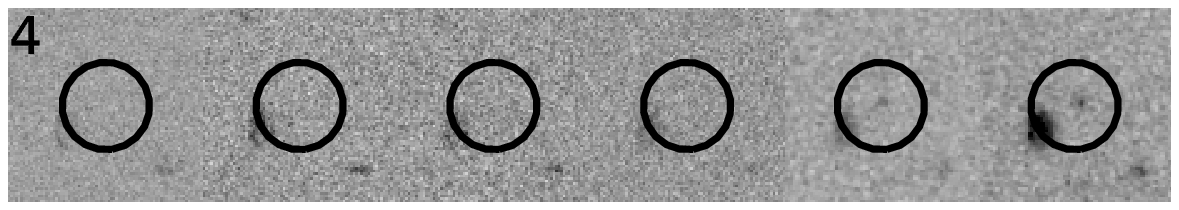} 
\includegraphics[width=32pc]{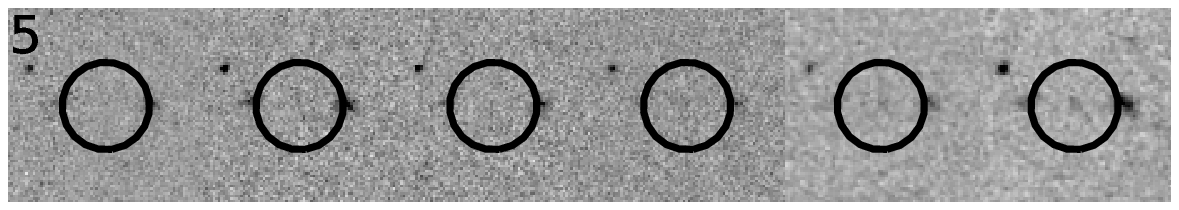} 
\includegraphics[width=32pc]{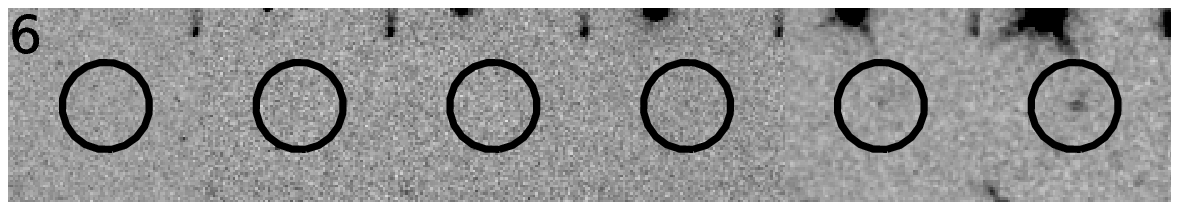}
\caption{Postage stamp images of our candidates. ACS $B$, $V$, $i$,
  $z$ images are shown alongside the WFC3 $Y$ and $J$-bands for
each of our candidates. Each image is $2.4$ arcsec across, with North up and East to the left.}
\end{figure*}

\section{Analysis}
\subsection{Candidate Selection}
Identification of candidates is achieved using the Lyman break
technique (e.g. Steidel et al. 1996), where a large colour
decrement is observed between filters either side of
 Lyman-$\alpha$  in the rest-frame of the galaxy. The flux
 decrement comes principally from the large 
integrated optical depth of the intervening absorbers
(the Lyman-$\alpha$ forest) .

In Figure 2 we illustrate how a colour cut of $(z-Y)>0.8$ is
effective at selecting sources at $z>6.5$. The principal caveat is
that for $(z-Y)<1.7$ we may include contamination from low-mass
galactic T and L dwarf stars and for $(z-Y)<1.0$ from evolved systems
at $z\approx 1.5$ where we pick up the Balmer/$4000{\rm \AA}$
break. 

These interlopers can be discriminated against using additional
colour information, and we use the $J$-band (F125W) images from this HST/WFC\,3 program together with existing ACS $BVi$-band imaging to achieve this.

Galactic L and T dwarf stars can be distinguished from high-$z$ sources by their position in $z-Y$-$Y-J$ colour plane (see Figure 5), as the low-mass stars typically possess redder $(Y-J)$ colours than the high-$z$ Lyman break galaxies. Lower redshift objects with strong Balmer breaks can however be more difficult to distinguish. Galaxies with strong Balmer breaks at $z\approx 1.5$ can mimic the $z-Y$ colours of $z'$-drop galaxies. These Balmer break galaxies have dominant post-starburst populations, of age $\ge 100$\,Myr. They will potentially be undetected at shorter wavelengths (the ACS filters) if they do not have significant ongoing star formation (i.e. the rest-UV is faint). Such objects will, however, have quite red $Y-J$ colours (above the Balmer break), and hence will lie in a different position in $z-Y$-$Y-J$  colour space than the Lyman break galaxies at $z\approx 7$ (see Figure~5). Figure~6b shows the photometry of an object in our sample which is well fit by a Balmer break spectrum at $z=1.3$ produced by an instantanous burst of age 3000\,Myr.

However, Balmer break galaxies can have bluer $Y-J$ colours if they have a more complicated star formation history, with a more recent or ongoing episode of star formation involving a small fraction of the stellar mass. At $z\approx 1.5$, these galaxies can lie in approximately the same position as $z\approx 7$ galaxies in the $z-Y$-$Y-J$ colour plane. Such objects will, however, also have a steep blue spectral slope shortward of break, and will protentially be detectable at shorter wavelengths (with a $V-Y$ colour of $\sim 1$ at $z\approx 1.5$). At our catalogue limit of $Y_{AB}=27.0$, these ``blue Balmer break'' galaxies will have $V$-band detections in the  ACS images of GOODS-South. Both a synthetic and observed example of such an object is shown in Figure~6c.

Our initial selection of objects brighter than $Y_{AB}=27.0$ and
redder than $(z-Y)_{AB}=0.8$ yielded 148 objects. Of these 37\% (55) are clearly spurious (e.g. diffraction spikes). Of the remaining 93 candidates all but 6 (6.5\%) have detections in the bluer $B$, $V$ or $i$ ACS bands at $>2\sigma$ (i.e. $B<29.14$, $V<29.16$, $i<28.45$). Of the 93 non-spurious objects, a further 8 (9\%) are detected in the ACS $i$-band but are undected at $>2\sigma$ in the $B$ \& $V$-bands. These objects are possibly $z\approx 6$ $i$-drop galaxies rather than $z\approx 7$ (indeed we recover galaxies CDFS-2290242079
and CDFS-2226643007 from the ACS GOODS $i$-drop catalog of Bouwens et al.\ 2006). 
Adopting a conservative approach, we do not include these $i$-band-detected sources in our final sample. Photometry of our 6 remaining $z$-drop candidates are presented in Table~1, and the ACS and WFC\,3 images are shown in Figure~4.

\begin{figure}
\centering
\includegraphics[width=18pc]{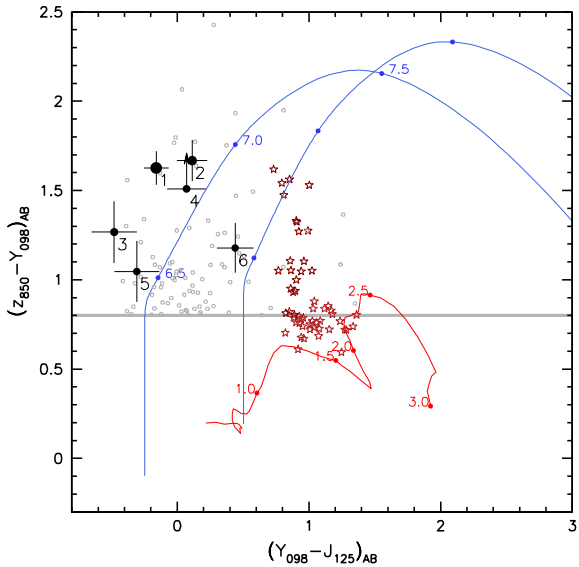} 
\caption{$(z'-Y)$ vs. $(Y-J)$ colour-colour diagram illustrating the
 differentiation of high-redshift star forming galaxies, low-mass L
 and T dwarf stars and lower redshift galaxies. The blue lines show the colour evolution of model starbursting galaxies with a
  ultraviolet spectral slopes of $\beta=-3$ (left) and $\beta=0.0$ (right)  (where
  $f_{\lambda}=\lambda^{\beta}$) damped below Lyman-$\alpha$ by a
  factor of $100$. The red line shows the colour evolution of an
  evolving stellar population that formed instantaneously at
  $z=10$. Red stars denote the observed colours of L and T dwarf stars
  (Chiu et al. 2006, Golimowski et al. 2004, Knapp et al. 2004). Objects matching our $z-Y>0.8$ selection criteria and which are also undected in the $BVi$ imaging are denoted by  the filled black circles and their 1$\sigma$ errors (or lower limits). Objects which meet the $z-Y$ criteria but are detected in one or more of the $BVi$ images are denoted by grey circles.}
\end{figure}

We also consider the available deep MIPS $24\mu m$ imaging of the field which covers all of our survey area. This MIPS band is sensitive to thermal re-emission from dust in moderate redshift sources (powered either by highly-obscured star formation or AGN activity), but we would not expect strong emission from $z\sim 7$ sources (rest-frame $3\,\mu$m). None of our 6 candidates are clearly detected (with $<10\,\mu$Jy at $2\,\sigma$), however objects 2,3, and 6 are confused by nearby sources. These objects nevertheless appear undetected, further supporting the assertion that our candidates are of high redshift origin.

\begin{figure}
\centering
\includegraphics[width=20pc]{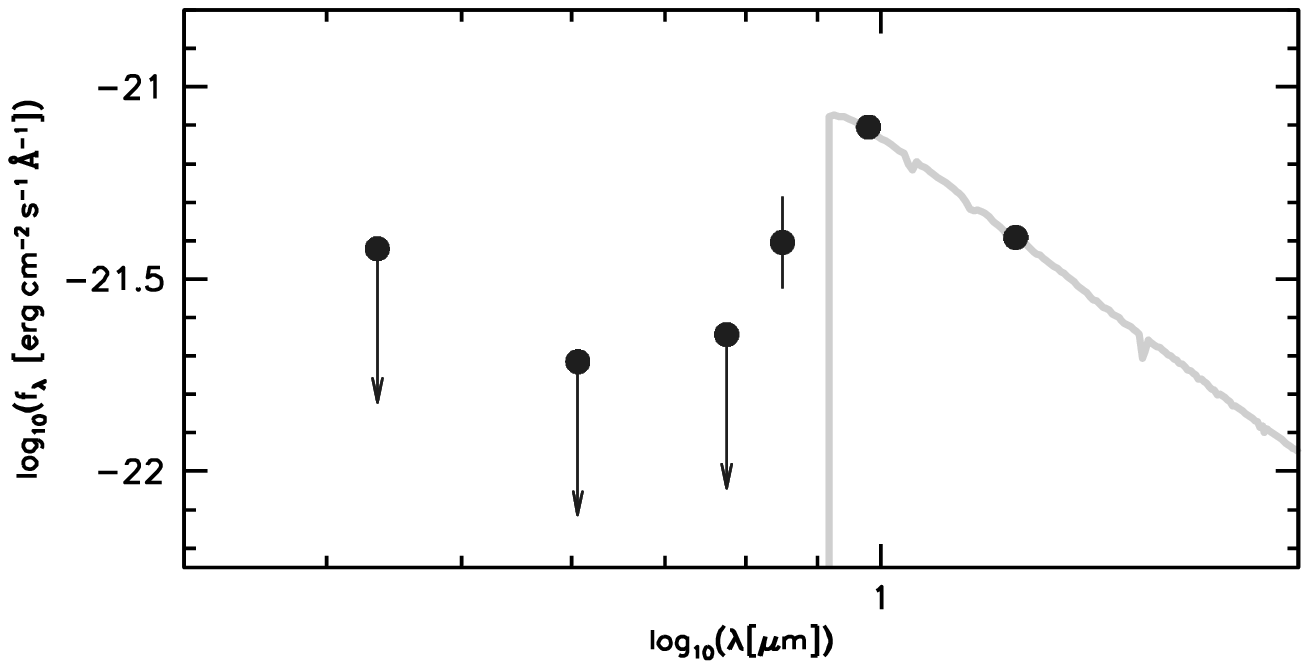} 
\includegraphics[width=20pc]{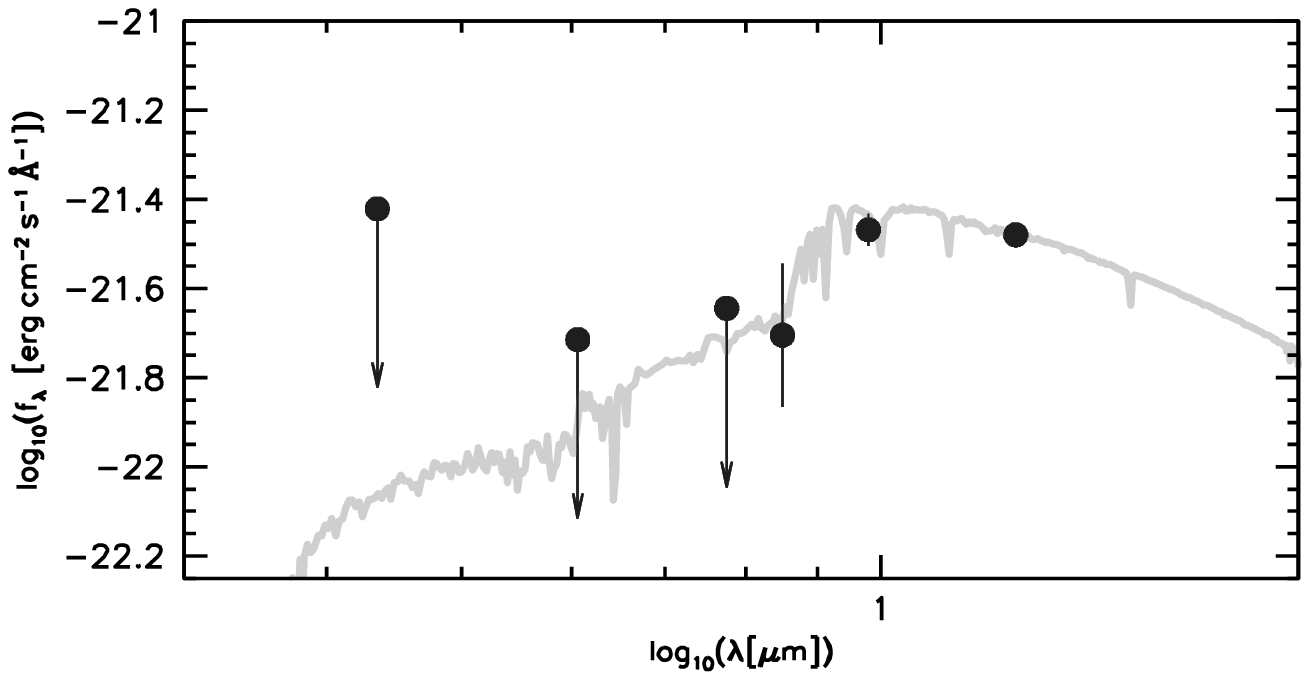} 
\includegraphics[width=20pc]{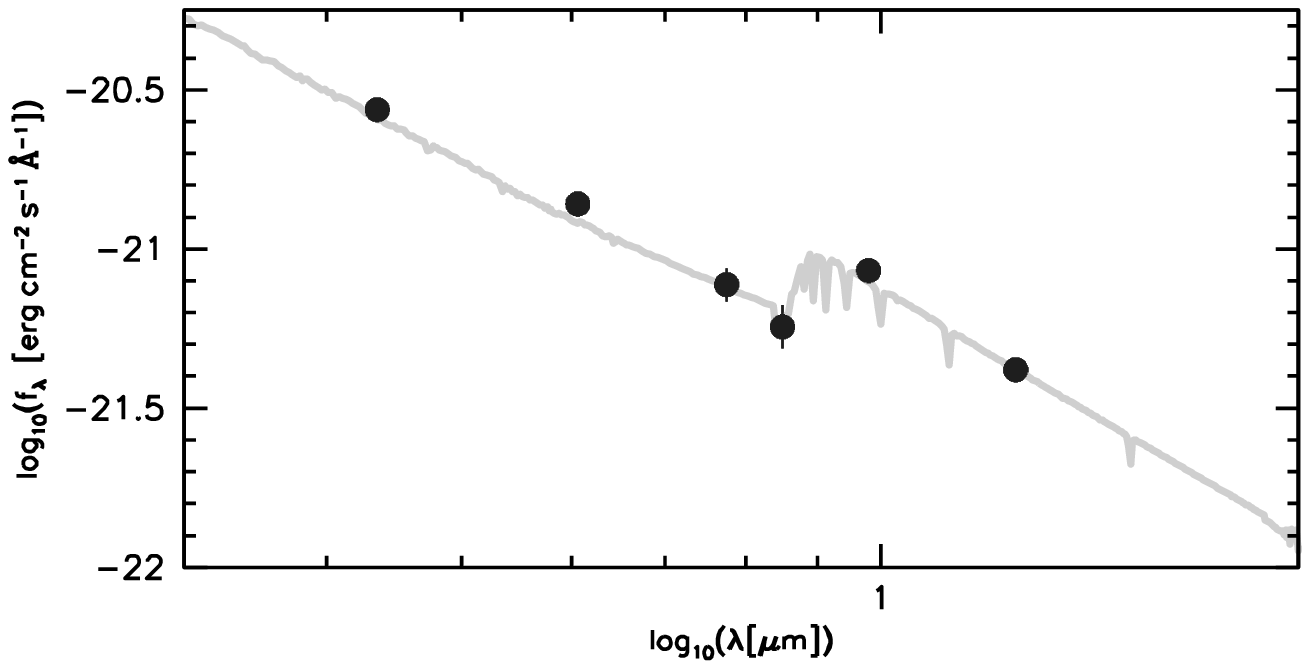} 
\caption{Example SEDs of objects meeting our $z-Y>0.8$ selection criteria. Top (a) - a candidate $z$-drop galaxy (points) and the synthetic SED of a $z\approx 7$ galaxy (line). Middle (b) - photometry of a possible Balmer break galaxy and the synthetic spectra of a galaxy composed of a single $3000\,$Myr population at $z=1.3$. Bottom (c) - an object detected in the ACS $BVi$-bands but which has a $J-Y$ colour consistent with high-$z$ galaxies. The grey line denotes the synthetic SED of galaxy comprised of two populations: a $200\,$Myr old instantaneous burst and a  $10\,$Myr old burst.}  
\end{figure}

There is a 5-year interval between the ACS images and the WFC3 observations,
which means that another possible source of contamination are
 transient objects (e.g. supernov\ae ), which might have brightened since
 the ACS data were taken. Indeed, the recent WFC\,3 images of the {\em Hubble}
 Ultra Deep Field reveal a probable supernova when compared with the deep ACS
 imaging in 2004 (Bunker et al.\ 2009; Oesch et al.\ 2009; Yan et al.\ 2009). 
 The WFC\,3 $Y$-band filter (F098M) used in the wide-field GOODS-South imaging
 presented here has greater overlap in wavelength with the $z$-band compared
 to the WFC\,3 F105W filter used for the {\em Hubble}
 Ultra Deep Field, which means that there is likely to be flux in the $z$-band for most
 $z\approx 7$ candidates. Indeed, in all but one case however each of the candidates are also
detected (faintly) in the ACS $z$ band images, implying they are
not extreme transient objects such as supernov\ae .
The one object (candidate 4) which is undetected in the $z$ band images has a 2$\sigma$
lower-limit of $(z-Y)_{AB}>1.5$. This, combined with its $Y-J$ colour,
suggests that it is still consistent with being a
high-$z$ star forming galaxy (Figure~5). Thus, we include this object in our
final candidate list.

\subsection{Number Density of Galaxies}

Observations of Lyman-break galaxies at $z>4$ have indicated significant evolution
of the luminosity function over cosmic time, in particular the discovery
that the number density of luminous star forming systems at $z\sim 6$ (the $i$-drops) is
much less than in the well-studied $U$-drop and $B$-drop population at $z\approx 3-4$ (e.g., Lehnert \& Bremer 2003; Stanway, Bunker
\& McMahon 2003; Bunker et al.\ 2004; Bouwens et al.\ 2004; Bouwens et al.\ 2006).

Recent results from the $z$-drops  indicate that this trend continues to $z=7-9$
(Bunker et al.\ 2009; Bouwens et al.\ 2009; Oesch et al.\ 2009; McLure et al.\ 2009; Hickey et al.\ 2009;
Ouchi et al.\ 2009). The new WFC3 data over several pointings allows us
to constrain the luminosity function around $Y_{AB}=26-27$, a key region around $L^*_{UV z\sim 6}$ with
too small numbers in the single Hubble Ultra Deep Field pointing (only one object along
with a probable supernova contaminant) but too faint
to be explored in the ground-based $Y$-band surveys.
We also note that the F098M  filter used in the data presented here does not go as red
as the F105W WFC\,3 filter used to survey the Hubble Ultra Deep Field (see Figure~1),
so we are biased towards slightly lower redshifts than $z=7$.

We have 6 candidates brighter than $Y_{AB}=27.0$ which are
 undetected in the ACS $B,V,i$ bands and with $(z-Y)_{AB}>0.8$. We survey 20\,arcmin$^{2}$
 over the 6 pointings (where we consider only the deepest region where all
 6 individual exposures per pointing overlap). Our surface densities are
 presented in Table~2 and Figure~7, where the error comes from the Poisson statistics. 

We now compare our observed surface density of $z\approx 7$ $z$-drop
candidates with that expected from a range of luminosity functions derived
for Lyman break galaxies at lower redshifts. We use the filter response
curves to assess which redshifts will satisfy our colour cut, assuming
 a simple model spectrum for these high-redshift star-forming
galaxy obeying a power law $f_{\lambda}\propto \lambda^{\beta}$ above
Lyman-$\alpha$ with a 99$\%$ flux decrement below due to the
Lyman-$\alpha$ forest during the Gunn-Peterson era (and no
flux below the 912\,\AA\ Lyman limit). For the UV
spectral slope we adopt a value of $\beta=-2$ consistent with the
values reported for high-redshift galaxies (Stanway, McMahon \& Bunker 2005), 
although we stress the number of objects is not
particularly sensitive to small variations in $\beta$. 

\begin{figure}
\centering
\includegraphics[width=18pc]{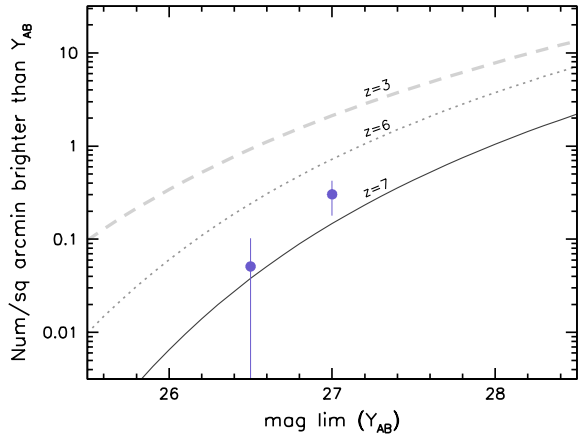} 
\caption{The cumulative surface density of $z$-drop galaxies ($y$-axis) brighter than a limiting
magnitude $Y_{lim}$ ($x$-axis). Our new results from the WFC\,3 imaging
of the GOODS-South field with the F098M filter are shown as the points. The dashed curve is the prediction using the UV luminosity function
derived for the $z=3$ Lyman break galaxies by Reddy et al.\ (2009); the dotted curve
is the luminosity function at $z=6$ from Bouwens et al.\ (2006) and the solidline is
the suggested $z=7$ luminosity function from Bouwens et al.\ (2008).}
\end{figure}

To compare our observations with predictions we consider three
UV luminosity functions, and note that for our assumed $\beta=-2.0$
(flat in $f_{\nu}$) there is no $k$-correction necessary to the AB magnitudes between
the various determinations of the UV continuum at 1300--1700\,\AA .
To assess evolution to $z\approx 3$ we use the recent Reddy et al.\ (2009)
determination of the Schechter luminosity function, which has a steep faint end slope
of $\alpha=1.73$, and characteristic luminosity of $M^*_{UV}=-20.97$
and number density of $\phi^*=0.00171\,{\rm Mpc}^{-3}$.
For our comparison to galaxies at $z\sim 6$, we adopt the Bouwens et al.\ (2006)
determination which has a similar faint end slope and $\phi^*=0.00206\,{\rm Mpc}^{-3}$
$M^*_{UV}=-20.25$. Finally, we consider the proposed $z\sim 7$ luminosity
function of Bouwens et al.\ (2008) with $\phi^*=0.0011\,{\rm Mpc}^{-3}$,
$M^*_{UV}=-19.8$ and a similar $\alpha$.
Our observations and predictions of the number densities of galaxies
are summarised in table 2 for a range of colour cuts $(z'-Y)\in
\{0.8,1.0,1.3\}$. 
Our observations suggest dramatic evolution between  $z\approx 6$ and $z\approx 7$,
with the observed number of $z$-drops  4$\sigma$ below the no-evolution
prediction. Our number density at $Y_{AB}<27$ is consistent with that obtained from our recent analysis (Bunker et al. 2009) of the UDF ($0.24\pm0.24\,{\rm arcmin}^{2}$). However it is roughly twice that suggested by the $z\approx 7$ luminosity function of Bouwens et al.\ (2008), although we emphasize that our statistical error is $\approx 40$ per cent and we estimate that the cosmic variance over our new data set will contribute a $30-40\%$ uncertainty (Trenti \& Stiavelli 2008). This is a significant improvement upon the $4.2$ arcmin$^2$ HUDF which contained only a single galaxy brighter than $Y=27$ (e.g. Bunker et al. 2009), precluding a reliable measure of the normalisation of the luminosity function owing to the large Poisson error ($\sim 100\%$) and cosmic variance ($\sim 50\%$) for sources in this magnitude range. 

The numbers of $z'$-drops discovered in this paper in the GOODS-South ERS-WFC3 observations are consistent with the deeper but smaller area
HUDF (Bunker et al.\ 2009 and Figure~7), accounting for differences in the shape of the $Y$-band filter bandpass. We consider the implications for reionization in Bunker et al.\ (2009): for our observed surface densities of $z'$-drops, the integrated star formation rate density (integrated down to $M_{UV}=-19.35 (AB)$, equivalent to $0.1\,L^*_{UV}$ at $z\approx 3$) is insufficient to achieve reionization unless the faint end slope is very steep ($\alpha \sim<-1.9$) and the escape fraction implausibly high $f_{\rm esc}>0.5$). However, the number of ionizing photons is determined by  the UV spectral slope (which is sensitive to the initial mass function, dust extinction, metallicity and star formation history). There are strong indications that the spectral slopes of our $z\approx 7$ Lyman-break galaxies are bluer than at lower redshift, and this evolution might make up the short-fall in ionizing photons.

The 6 $z$-drop galaxies presented here have
star formation rates of $5-10\,M_{\odot}\,{\rm yr}^{-1}$,
assuming they are are at $z=7$.
For a rest-frame equivalent width of 10\,\AA\ for Lyman-$\alpha$ (typical of
Lyman break galaxies at modest redshift, $z\approx 3-4$, Steidel
et al.\ 1999) this would
correspond to a line flux of $1-2\times 10^{-18}\,{\rm ergs\,cm^2\,s^{-1}}$, and for larger
equivalent widths of 30\,\AA , which may be more typical of the $z>5$ population
(Stanway et al.\ 2004; Vanzella et al.\ 2009) these galaxies may exhibit Lyman-$\alpha$
line emission fluxes $3-6\times 10^{-18}\,{\rm ergs\,cm^2\,s^{-1}}$. These fluxes are within
the capability of red-sensitive optical spectrographs on 8--10\,m telescopes with long integrations. Given the large fraction of $z=5-6$ dropouts in the same
luminosity range that show such strong Ly-$\alpha$ emission (Stark et al. 2010),
this sample should enable us to test whether the prevalence of Ly-$\alpha$
emitters declines over $6<z<7$, as expected given observations of narrowband
selected Ly-$\alpha$ emitters (e.g. Ota et al. 2008).

\begin{table*}
\begin{tabular}{cccccc}
Colour cut & Limiting magnitude & Observed & $z=3$ pred.$^{a}$ & $z=6$ pred.$^{b}$ & $z=7$ pred.$^{c}$ \\
\hline\hline
0.8 & 26.5 & 0.051$\pm$0.051 &0.926  &0.241 &0.038 \\
0.8 & 27.0 & 0.303$\pm$0.123 &2.114  &0.722 &0.147 \\
\hline
1.0 & 26.5 & 0.051$\pm$0.051 &0.712  &0.173 &0.026 \\
1.0 & 27.0 & 0.303$\pm$0.123 &1.678  &0.546 &0.106 \\
\hline
1.3 & 26.5 & 0.051$\pm$0.051 &0.388 &0.079 &0.010 \\
1.3 & 27.0 & 0.152$\pm$0.087 &0.987 &0.284 &0.049 \\
\end{tabular}
\caption{Observed and predicted surface densities of $z$-drop
  assuming different colour selections and luminosity
  functions. $^{a}$ assuming the Reddy et al.\ (2009) $z=3$ UV luminosity
function.  $^{b}$ assuming the Bouwens et al.\ (2006) $z=6$ UV luminosity
function. $^{c}$ assuming the Bouwens et al.\ (2008) $z=7$ UV luminosity
function.}
\end{table*}

\section{Conclusions}

In this work we have searched for star-forming galaxies at
$z\approx 7$ utilising the Lyman-break technique on newly acquired
F098M $Y$-band images from WFC3 on the {\em Hubble} Space
Telescope. Through the comparison of these images to existing {\em
  Hubble} ACS F850LP $z$-band images we identified objects with red
colours, $(z-Y)>0.8$, indicative of a break in the spectrum. We explore
an area five times larger than the recent WFC\,3 imaging of the {\em Hubble}
Ultra Deep Field. The new wider-field data in GOODS-South probes down
to $Y_{AB}=27$, equivalent to $\approx L^*_{UV}$ at $z=7$.

Using additional imaging (ACS $B$, $B$, $i$-bands from GOODS v2.0, and WFC3
F125W $J$-band) we removed contaminating objects which were either
detected in the bluer ACS bands or with observed near-infrared colours
inconsistent with high-redshift star forming galaxies.

This selection criteria left 6 candidates down to a limiting magnitude $Y_{AB}<27.0$ (equivalent to a star formation rate of $4.5\,{\rm
  M_{\odot}\,yr^{-1}}$ at $z=7$) of which all but one were detected in the
$z$-band (and are thus likely not to be transients). This implies a
surface density of objects brighter than $Y_{\rm AB}=27.0$ of 0.30$\pm$0.12$\, {\rm arcmin}^{2}$; a value smaller than both the prediction based on the
observed $z\approx 3$ and $z\approx 6$ luminosity functions,
suggesting continued evolution of the LF beyond $z=6$.

Knowledge of the surface density of dropouts in this
magnitude range is crucial in constraining the luminosity function, as current estimates of the $z\approx 7$ luminosity
function indicate that an $L_{*}$ galaxy has a magnitude of $Y\sim 27$ (e.g. Bouwens et al. 2008). Determining the UV luminosity
function is crucial in order to address whether star-forming galaxies could plausibly have provided the Lyman continuum photons
necessary to reionize the Universe.

Given the difficulty that ground-based surveys face in probing faintward of $Y=26$, larger-area surveys with {\em HST}/WFC3 ($\sim 100$ arcmin$^2$) probing to similar depths ($Y(6\sigma)\sim 27.2$) as the dataset presented in this paper will deliver a factor of 2 improvement in the poisson uncertainty and a significant decrease in the uncertainty due to cosmic variance.

\subsection*{Acknowledgements}

We would like to thank the annoyamous referee for their timely response and useful suggestions. Additionally we thank Mark Lacy for comments given during the construction of the manuscript, and Richard McMahon, Jim Dunlop, Ross McLure, Masami Ouchi, Bahram Mobasher and Michele Cirasuolo for many useful discussions about Lyman break galaxies at high redshift. Based on observations made with the NASA/ESA Hubble Space Telescope,
obtained from the Data Archive at the Space Telescope Science Institute, which is operated by the Association
of Universities for Research in Astronomy, Inc., under NASA contract
NAS 5-26555. These observations are associated with programme \#GO/DD-11359. We are grateful to
the  WFC\,3 Science Oversight Committee for making their Early Release Science observations public.
SW, DS \& ES acknowledges funding from the U.K.\ Science and Technology Facilities Council. MJJ acknowledges the support of a RCUK fellowship. SL is supported by the Marie Curie Initial Training Network ELIXIR of the European Commission under contract PITN-GA-2008-214227.

\bsp

\end{document}